\documentclass{article}

\usepackage{arxiv}

\usepackage[utf8]{inputenc} 
\usepackage[T1]{fontenc}    
\usepackage{hyperref}       
\usepackage{url}            
\usepackage{booktabs}       
\usepackage{amsfonts}       
\usepackage{nicefrac}       
\usepackage{microtype}      
\usepackage{xcolor}         

\usepackage{natbib}
\usepackage{graphicx}
\usepackage{subcaption}
\usepackage{amsmath}
\usepackage{amssymb}
\usepackage{fixltx2e}
\title{Safe-FinRL: A Low Bias and Variance Deep Reinforcement Learning Implementation for High-Freq Stock Trading}

\author{%
   Zitao Song
    \\
  School of Data Science\\
  The Chinese University of Hong Kong (Shenzhen)\\
  Shenzhen, Guangdong 518172 \\
  \texttt{zitaosong@link.cuhk.edu.cn} \\
   \And
   Xuyang Jin \\
   School of Data Science\\
  The Chinese University of Hong Kong (Shenzhen)\\
  Shenzhen, Guangdong 518172 \\
   \AND
   Chenliang Li \\
   School of Data Science\\
  The Chinese University of Hong Kong (Shenzhen)\\
  Shenzhen, Guangdong 518172 \\
}

\date{}

\renewcommand{\shorttitle}{Safe-FinRL}

\begin{document}
\maketitle

\begin{abstract}
In recent years, many practitioners in quantitative finance have attempted to use Deep Reinforcement Learning (DRL) to build better quantitative trading (QT) strategies. Nevertheless, many existing studies fail to address several serious challenges, such as the non-stationary financial environment and the bias and variance trade-off when applying DRL in the real financial market. In this work, we proposed Safe-FinRL, a novel DRL-based high-freq stock trading strategy enhanced by the near-stationary financial environment and low bias and variance estimation. Our main contributions are twofold: firstly, we separate the long financial time series into the near-stationary short environment; secondly, we implement Trace-SAC in the near-stationary financial environment by incorporating the general retrace operator into the Soft Actor-Critic. Extensive experiments on the cryptocurrency market have demonstrated that Safe-FinRL has provided a stable value estimation and a steady policy improvement and reduced bias and variance significantly in the near-stationary financial environment. 
\end{abstract}


\section{Introduction}
Stock trading is considered one of the most challenging decision processes due to its heterogeneous and volatile nature. In recent years, many practitioners in quantitative finance have attempted to use Deep Reinforcement Learning (DRL) to extract multi-aspect characteristics from complex financial signals and make decisions to sell and buy accordingly. Specifically, in DRL, the stock trading problem is modeled as a Markov Decision Process (MDP) problem. This training process involves observing stock price changes, taking actions and reward calculations, and letting the trading agents adjust their policies accordingly. Through integrating many complex financial features, the DRL trading agent builds multi-factor models and provides algorithmic trading strategies that are difficult for human traders.

In recent research, DRL has been applied to construct various trading strategies, including single stock trading, multiple stock trading, portfolio allocation, and high-frequency trading. \citet{liu2020finrl} introduce a DRL library called FinRL and implement several bench-marking DRL algorithms to construct profitable trading strategies in S\&P 500 index. \citet{jiang2017deep} apply Deterministic Policy Gradient (DPG) to the portfolio allocation problem of cryptocurrency. \citet{wang2021deep} develop a hierarchy DRL that contains a high-level RL to control the portfolio weights among different days and a low-level RL to generate the selling price and quantities within a single day. \citet{fang2021universal} implement a DRL Oracle by distilling actions from future stock information. Despite the fruitful studies above, applying DRL to the real-life financial markets still faces several challenges and current research fails to address those elegantly.

The first challenge is that \textit{No man ever steps in the same river twice}. Since in the real-life heterogeneous financial market, there always exists a huge discrepancy in stock prices between different periods. Although it is an appealing approach to use some nonlinear function approximators such as neural networks to estimate the value function in the financial market, a stable value function estimation and a steady policy improvement are extremely difficult because of the incoming non-stationary financial data. Consequently, when using DRL in the financial market, the value or policy function will always fail to converge given the non-stationarity between episodes in the long-time-range financial environment. Existing research \citep{liu2020finrl,jiang2017deep} also fails to provide such an empirical or theoretical convergence analysis for their DRL model training in a non-stationary environment.

The second challenge is \textit{The trade-off of bias and variance in off-policy learning in the financial environment}. In traditional RL, there always exists a trade-off between using the Temporal Difference (high bias and low variance) and the Monte Carlo (high bias and low variance) to estimate the value function. Due to the noisy and volatile nature of the financial market, this bias and variance trade-off has become even more critical when training our DRL trading agent. In practice, the single-step return in the financial market is highly volatile. Thus, the value function estimated by minimizing the single-step bellman error tends to have a huge bias. On the other hand, an extended multi-step DRL in an off-policy learning setting will lead to both the increasing variance and the distribution shift between the target policy and the behavior policy. Consequently, it is significant for us to consider the bias and variance trade-off and implement off-policy multi-step DRL properly in the financial market.

In this study, we propose Safe-FinRL, which mainly focuses on these two challenges of the applied DRL algorithms in the high-frequency stock trading market. In particular, we have provided a \textit{Safe} way to train off-policy FinRL agents on the financial market with reduced bias and variance and have also considered the distribution shift in off-policy learning. Specifically, our total contributions are threefold. Firstly, we attempt to minimize the negative effect of the non-stationary financial environment by separating long time financial periods into four independent small parts. Within this short time slot, we assume the financial environment to be near-stationary. Secondly, we implement the single-step Soft Actor-Critic (SAC) \citep{haarnoja2018soft} on the four independent environments as benchmarks. Finally, we realize Safe-FinRL by Trace-SAC, which incorporates the multi-step TD learning to the plain SAC and uses different traces to correct the distribution shift in off-policy learning. Extensive experiments on the artificial high-frequency trading environment have shown that Safe-FinRL has lower bias and variance than the plain SAC when estimating the value function. In addition, Safe-FinRL also has obtained positive revenues in all of the proposed four environments and outperformed the market in two of them.

\section{Related Work}
Here we review some related works on the relevant SOTA RL algorithms, recent research on the bias and variance reduction in the off-policy RL algorithms, and existing implementations of DRL in finance.
\subsection{Stated-of-the-Art RL Algorithms}
Generally, DRL algorithms can be categorized into three approaches: value-based algorithm, policy-based algorithm, and actor-critic algorithm. For value-based algorithms, they mainly focus on the discrete action space control problem and use the bellman equation to update. Among them, Deep Q-Network (DQN) \citep{mnih2015human,mnih2013playing} adapts Q-learning and eplison-greedy technique, and it has been successfully applied to playing video games. For policy-based algorithm, Proximal Policy Optimization (PPO) \citep{schulman2017proximal} and Trust Region Policy Optimization (TRPO) \citep{schulman2015trust} improve the vanilla Policy Gradient \citep{sutton1999policy} by introducing Importance Sampling (IS) to correct the distribution shift and Generalized Advantage Estimation (GAE) \citep{schulman2015high} to obtain stable and steady improvement in some 3D-locomotion tasks. For actor-critic based algorithm, Deep Deterministic Policy Gradient (DDPG) \citep{lillicrap2015continuous} and Twin-Delayed DDPG (TD3) \citep{fujimoto2018addressing} uses value and policy function as two function approximators and optimize them simultaneously. Soft Actor-Critic (SAC) \citep{haarnoja2018soft} introduces an additional entropy term to its objective function to facilitate exploration based on DDPG and TD3. 

\subsection{Bias and Variance Reduction in Off-policy Learning}

In many real-world DRL applications, interaction with the environment is costly, and we want to reuse the sample trajectories collected before. This necessitates the use of \textit{off-policy} DRL methods, algorithms that can learn from previous experience collected by possibly unknown behavior policies. In the actor-critic algorithm, its value function is learned by iteratively solving the Bellman Equation, which is inherently off-policy and independent of any underlying data distribution \citep{nachum2019algaedice}. Nevertheless, when multi-step returns are used to control bias in off-policy learning, the distribution shift between target policy and behavior policy will occur inevitably. Importantly, this discrepancy can lead to complex, non-convergent behavior of value function in the algorithm \citep{kozuno2021revisiting}.

Currently, there are two main approaches two handle this distribution shift. One approach is to manually correct the shift by setting `trace values' to the multi-step bellman equations. Among them, Retrace \citep{munos2016safe}, Tree-Backup (TBL) \citep{precup2000eligibility} are \textit{conservative} methods that the convergence of value function is guaranteed no matter what behavior policy is used. By contrast, C-trace \citep{rowland2020adaptive}, Peng's Q($\lambda$) (PQL) \citep{peng1994incremental}, and uncorrected n-step return are \textit{non-conservative} methods that do not damp the trace value or truncate trajectories, and thus do not guarantee generic convergence. Importantly, the `trace' approach can be extended to the actor-critic settings easily. Another approach, DItribution Correction Estimation (DICE) families \citep{nachum2019algaedice,nachum2019dualdice}, are alternatives to the policy gradient and value-based methods, which directly give the estimation of the distribution shift by introducing a new formulation of max-return optimization.

\subsection{DRL in Stock Trading}
Among the recent research on DRL applications for stock trading, many of them follow the model-free off-policy DRL settings. \citet{jiang2017deep} apply Deterministic Policy Gradient (DPG) to the portfolio allocation problem of cryptocurrency. \citet{liang2018adversarial} model stock trading as a continuous control problem and implement DDPG and PPO for portfolio management in China's stock market. \citet{liu2020finrl} introduce a DRL library called FinRL and implement several benchmarking DRL algorithms to construct profitable trading strategies in S\&P 500 index. \citet{wang2021deep} develop a hierarchy DRL that contains a high-level RL to control the portfolio weights and a low-level RL to decide the specific selling price and quantities. \citet{fang2021universal} implement an RL Oracle by distilling actions from future stock information. Despite many of them claiming to be profitable, very few of them have customized their DRL models to the heterogeneous and volatile financial environment.

\section{Problem Formulation}

\subsection{Markov Decision Process Formulation}
Under the assumption that the pricing of one financial asset only depends on its previous prices within a certain range, we model the stock trading process as a Markovian Process. Specifically, when trading an asset in the real financial market, both the price and the decision of selling or buying the asset at time step $t$ are singly determined by their previous information, and the return of that decision is singly generated by the price changing from time $t$ to $t+1$. Consequently, the financial task of trading assets can be modeled as a Markov Decision Process (MDP). The detailed definitions of the state space, action space, and reward function in our model are shown below.

\textbf{State Space} $\mathcal{S}$. Normally, state space of financial assets only contain their previous pricing information $\mathbf{p_{t}}$. In our work, we extend the state space $\mathbf{s_{t}}$ in three direction. First, we add over 118 technical features $\mathbf{q_{t}}$ for each asset in the high-frequency trading market at time $t$ to the original pricing information. Second, we incorporate the remaining balance $\mathbf{b_{t}}$ and holding shares for each asset $\mathbf{h_{t}}$ as a part of state space at time $t$. Finally, instead of looking at single time $t$'s information, we use a look-back window with length $l$ (default:3) and enlarge $\mathbf{s_{t}}$ by stacking previous information from $t-l$ to $t$. In summary, $\mathbf{s_{t}}$ contains three parts,
\begin{itemize}
    \item Balance $\mathbf{b_{t}}$: the amount of money left in the account at the time step $t$
    \item Holding Shares $\mathbf{s_{t}}$: the amount of holding shares for each asset at time step $t$.
    \item Features $\{ \mathbf{q_{t-l:t}}, \mathbf{p_{t-l:t}}\}$: technical and price features for each asset from time step $t-l$ to $t$.
\end{itemize}

\textbf{Action Space} $\mathcal{A}$. The action space of trading one asset is formulated by a continuous one-dimension vector $\mathbf{a_{t}} \in [-h\_max,h\_max]$ that represents the confidence of longing or shorting that asset, where $h\_max$ is the upper and lower bound of the action (default:0.1). When $\mathbf{a_{t}}$ closes to $h\_max$, it represents a strong signal to buy that asset. On the contrary, $\mathbf{a_{t}}$ closes to $-h\_max$ represents high confidence to sell that asset. The amount of selling or buying shares are calculated by $ \lfloor \frac{\mathbf{a_{t}}}{u}\rfloor \cdot u$, where $u$ is the minimum trading unit in the high frequency trading market. 

\textbf{Reward} $\mathcal{R}(s,a,s')$. Consider $s'$ is the new arriving state when action $a$ is taken at state $s$. Reward $r(s,a,s')$ is formulated by $\log(\frac{v'}{v})$, where $v'$ and $v$ is the total amount of wealth at state $s'$ and state $s$. Specifically, at time t, $v_{t} = b_{t} + p_{t} \cdot a_{t} \cdot \text{sgn}(-a_{t}) \cdot (1 + \text{sgn}(a_{t})\cdot c)$, where $c$ is the commission fee and $p_{t}$ is the sell/bid price when $a_{t} < 0$ and $p_{t}$ is the buy/ask price when $a_{t} > 0$.

\subsection{Artificial Financial Environment}

Since in the financial market, the transition probability from state $s_{t}$ to state $s_{t+1}$ can not be explicitly modeled, based on OpenAI Gym, we build an artificial trading environment with real-life market data. Unlike trading in the real world, we have loosened restrictions on positions, assuming that the agent can buy and sell any amount of shares in that trading environment. Finally, the trading robot can observe states, take actions of long and short, receive rewards, and have its strategy adjusted accordingly through iteratively interacting with the built artificial trading environment.

\subsection{Dataset}

The raw data of the real life market consists of snapshots of order book data and tick-by-tick trade data of the btc-usdt swap of Binance. We aggregate the raw data by minute and formulate 118 meaningful predictors as features. All features are standardized by z-score to ensure the stability of the model. We also designed a trade book that contains bid prices and ask prices at the end of each minute for simulated trading since the spread is crucial in high-frequency trading. The data is for 20 days, from \textbf{March 10, 2022} to \textbf{March 29, 2022}. After processing, the total sample size is around 28800.  

\section{Methodology}

\subsection{Baselines}
In this work, we choose the single-step Soft Actor-Critic (SAC) \citep{haarnoja2018soft} as our benchmark algorithm. This is because through generating a stochastic policy and maximizing an adjustable entropy term in its objective function, SAC can capture multiple modes of near-optimal behavior and thus has superior exploration ability over other deterministic policy algorithms such us DDPG \citep{lillicrap2015continuous} and TD3 \citep{fujimoto2018addressing}. In our experiment, we find that setting the weight of the entropy term in SAC to be constant will help convergence, so we choose to use SAC with an unadjustable entropy term as the benchmark. Moreover, we also use the market value itself as one of our benchmarks.

\subsection{Environment Separation}
\begin{figure}[tp]
    \centering
    \includegraphics[width=\textwidth]{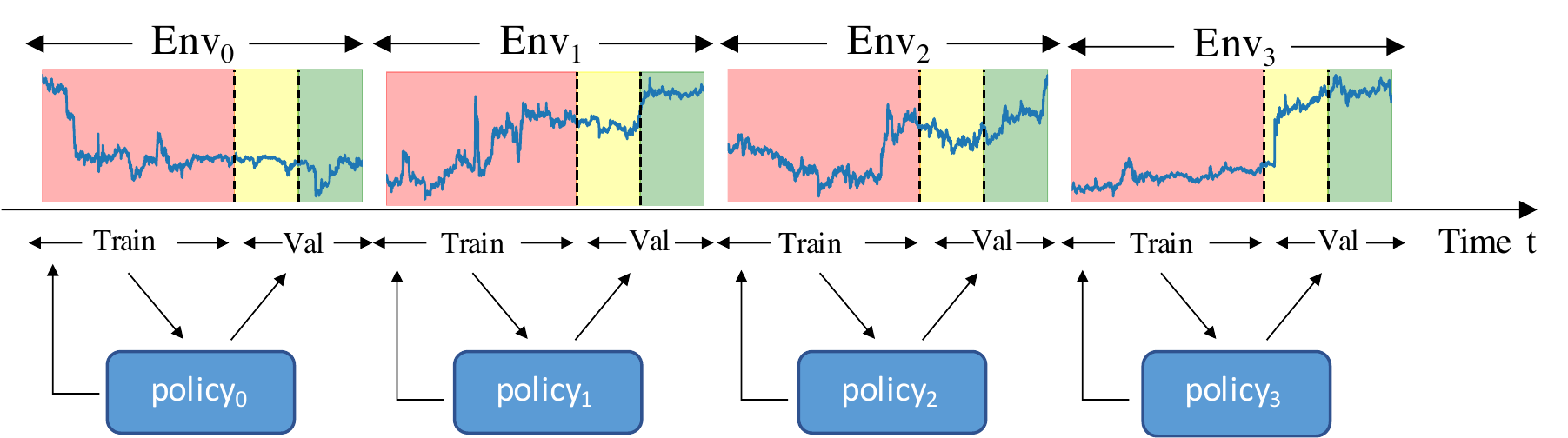}
    \caption{Environment Separation of High Freq Trading Environment}
    \label{fig:ev1}
\end{figure}

 Here, we separate the twenty days of trading data into four independent near-stationary environments to reduce the negative effect of a non-stationary stock trading environment. Each separated environment contains five days of trading data, of which three days are used to train, and the left two days are used to do validation and test. In Figure \ref{fig:ev1}, we present the detailed process. From the figure, we can observe that the consecutive four environments indeed are highly heterogeneous. In Env\textsubscript{0}, the stock price displays an obvious descending trend while Env\textsubscript{3} witnesses a remarkable stock price spike. On the contrary, the stock price trend within each environment is more stable and predictable than the highly non-stationary trend between separated environments.
 
 During training, we let four agents interact with these four environments independently and generate four different policies accordingly. After training five episodes on each environment, we will test the policy on the validation data (environment).

\subsection{Trace-SAC}
We first briefly review the definitions of policy and value functions in RL and build up the necessary knowledge for multi-step TD learning. Next, we introduce the off-policy multi-step TD learning based on the General Retrace operator \citep{munos2016safe}. Finally, we explain the proposed Trace-SAC algorithms.

\textbf{Single-Step Estimation}. We consider the standard off-policy RL setup where an agent interacts with an environment, generating a sequence of state-action-reward-state tuples $(s_{t},a_{t},r_{t}, s_{t+1})$, where $s_{t}, s_{t+1} \in \mathcal{S}$ and $a_{t} \in \mathcal{A}$ which is sampled from some policy $\pi_{k} : \mathcal{S} \to [0,1]$. Here we define the discounted cumulative return start at time $t$ as $G_{t} := \sum_{t'=t}^{\infty}\gamma^{t'-t}r_{t'} $. The value function and Q-function at time $t$ are defined by $V^{\pi_{k}}(s_{t}) :=\mathbb{E}_{\pi_{k}}[G_{t}|s_{t}]$ and $Q^{\pi_{k}}(s_{t},a_{t}):=\mathbb{E}_{\pi_{k}}[G_{t}|s_{t},a_{t}]$, respectively, where the expectation over $\pi_{k}$ means $a_{t} \sim \pi_{k}(\cdot | s_{t})$. 

Since large continuous domains require us to design practical function approximators for both policy and value function, we assume Q-function $Q_{\theta}: \mathcal{S} \times \mathcal{A} \to \mathbb{R}$ is parameterized by $\theta$ and policy function $f_{\phi}:\mathcal{S} \to \mathcal{A}$ is parameterized by $\phi$. Based on that, the single step bellman operator $Q_{k+1}:=\mathcal{B}^{\pi_{k}}Q_{k}$ and bellman optimal operator $Q_{k+1}^{*}:=\mathcal{B}^{*}Q_{k}^{*}$ are given by:

\begin{align}
\mathcal{B}^{\pi_{k}}Q_{k}(s_{t},a_{t}) &:= r_{t} + \gamma \mathbb{E}_{s_{t+1} \sim \rho_{\pi_{k}}}[Q_{k}(s_{t+1},f_{\phi}(s_{t+1}))] \\
\mathcal{B}^{*}Q_{k}^{*}(s_{t}, a_{t}) & := r_{t} + \gamma \mathbb{E}_{s_{t+1} \sim \rho_{\pi_{k}}}[\text{max}_{a'}Q^{*}_{k}(s_{t+1},a')]
\end{align}
where $\rho_{\pi_{k}}(s_{t})$ is the state marginal of the trajectory distribution induced by policy $\pi_{k}$. Next, the Temporal Difference (TD) or single-step bellman error is obtained by $\delta_{t} := \mathcal{B}^{\pi_{k}}Q_{k}(s_{t},a_{t}) - Q_{k}(s_{t},a_{t})$ and $\delta_{t}^{*} := \mathcal{B}^{*}Q_{k}(s_{t},a_{t})- Q_{k}(s_{t},a_{t})$. Importantly, the first obtained $\delta_{t}$ is often used in the value function approximation in some actor-critic algorithms, such us DDPG \citep{lillicrap2015continuous} while the second $\delta_{t}^{*}$ is used in the standard Q-learning algorithm \citep{watkins1992q}. Finally, by using Stochastic Approximation \citep{robbins1951stochastic}, we approximate the Q-function by iteratively calculating:
\begin{align}
    Q_{k+1}(s_{t},a_{t})  :&= (1-\alpha_{k}) Q_{k}(s_{t},a_{t}) + \alpha_{k}(\mathcal{B}^{\pi_{k}}Q_{k}(s_{t},a_{t})) \\
    & = Q_{k}(s_{t},a_{t}) + \alpha_{k} \delta_{t}
\label{eq:single}
\end{align}
where $\alpha_{k} \to 0$ as $k \to \infty$.

\textbf{General Retrace}. \citet{munos2016safe} extend the single-step value function estimation in \eqref{eq:single} and introduce a general retrace operator to estimate the value function using both off-policy and multi-step TD learning, from which the distribution correction can be obtained by singly changing the trace value. Specifically, the general retrace algorithm updates its Q-function by $Q_{k+1} := \mathcal{R}_{\lambda}^{c_{k},\mu_{k},\pi_{k}}Q_{k}$, where $(c_{k})_{k\ge0}$ is a sequence of non-negative functions over $\mathcal{S} \times \mathcal{A}$. $(\mu_{k})_{k \ge 0}$ is an arbitrary sequence of behavior policies, and $(\pi_{k})_{k \ge 0}$ is a sequence of target policies that depends on an algorithm. Given a trajectory $(s_{t},a_{t},r_{t},s_{t+1})_{t\ge 0}$ collected under $\mu_{k}$, the general retrace algorithm $ Q_{k+1}(s_{t},a_{t}):=(\mathcal{R}_{\lambda}^{c_{k},\mu_{k},\pi_{k}}Q_{k})(s_{t},a_{t})$ can be written as:
\begin{equation}
     Q_{k+1}(s_{t},a_{t}) : =Q_{k}(s_{t},a_{t}) + \sum_{t'=t}^{\infty}\Big(\prod_{u=1}^{t'}c_{k}(s_{u},a_{u})\Big)\gamma^{t'}\lambda^{t'}\delta_{t'},
\label{retrace}
\end{equation}
where $\prod_{u=1}^{t}c_{k}(s_{u},a_{u}):=1$ and $\delta_{t} = \mathcal{B}^{\mu_{k}}Q_{k}(s_{t},a_{t})-Q_{k}(s_{t},a_{t})$.

\textbf{Trace-SAC}. In the single-step Soft Actor Critic \citep{haarnoja2018soft}, its target is adjusted to maximize a soft Q-function $Q^{soft} (s_{t}, a_{t}):= r_{t} + \mathbb{E}_{(s_{t+1},\cdots)\sim \rho_{\pi}}[\sum_{l=1}^{\infty} \gamma^{l}(r_{t+l}+\alpha \mathcal{H}(\pi^{*}_{\text{MaxEnt}}(\cdot|s_{t+l})))]$, where $\mathcal{H}(\cdot)$ is the Shannon entropy and the details of $\pi^{*}_{\text{MaxEnt}}$ can be found in \citep{haarnoja2017reinforcement}. Based on that, the soft policy evaluation is defined as $Q_{k+1}^{soft} := \mathcal{T^{\pi}}Q_{k}^{soft}$, where $Q^{soft}: \mathcal{S} \times \mathcal{A} \to \mathbb{R}$ and $ \mathcal{T}^{\pi}$ is a modified bellman backup operator. Given a single step tuple $(s_{t},a_{t},r_{t},s_{t+1})$, the modified bellman backup operator is given by:
\begin{equation}
    \mathcal{T}^{\pi}Q_{k}^{soft}(s_{t},a_{t}) := r_{t} + \gamma \mathbb{E}_{s_{t+1} \sim \rho_{\pi_{t}}, a_{t} \sim \pi_{t}}[Q_{k}^{soft}(s_{t+1},f_{\phi}(s_{t+1})) - \log \pi_{t}(f_{\phi}(s_{t+1}))]
\end{equation}
where $\rho_{\pi_{k}}(s_{t})$ is the state marginal of the trajectory distribution induced by policy $\pi_{t}$. According to \citet{haarnoja2018soft}, the soft bellman backup operator $\mathcal{T}^{\pi}$ shares a similar convergence property as $\mathcal{B}^{\pi}$. Specifically, the sequence $Q_{k}^{soft}$ will converge to the soft Q-value of policy $\pi$ as $k \to \infty$. This motivates us incorporate SAC into the General Retrace algorithm.

In the Trace-SAC we proposed, we use the convergence property of soft bellman backup operator $\mathcal{T}^{\pi}$ and integrate SAC into the General Retrace algorithm. In practice, the Trace-SAC updates its Q-function by $Q_{k+1}^{soft}:=\mathcal{R'}_{\lambda}^{c_{k},\mu_{k},\pi_{k}}Q_{k}^{soft}$, where $(c_{k})_{k\ge0}$ is a sequence of non-negative functions over $\mathcal{S} \times \mathcal{A}$. $(\mu_{k})_{k \ge 0}$ is an arbitrary sequence of behavior policies, and $(\pi_{k})_{k \ge 0}$ is a sequence of target policies that depends on an algorithm. Given a trajectory $(s_{t},a_{t},r_{t},s_{t+1})_{t\ge 0}$ collected under $\mu_{k}$, the soft value function estimation for Trace-SAC can be written as:
\begin{equation}
     Q_{k+1}^{soft}(s_{t},a_{t}):=Q_{k}^{soft}(s_{t},a_{t}) + \sum_{t'=t}^{T}\Big(\prod_{u=1}^{t'}c_{k}(s_{u},a_{u})\Big)\gamma^{t'}\lambda^{t'}\delta_{t'},
\label{retrace}
\end{equation}

where $\prod_{u=1}^{t}c_{k}(s_{u},a_{u}):=1$ and $\delta_{t'} = \mathcal{T}^{\mu_{k}}Q_{k}^{soft}(s_{t'},a_{t'})-Q_{k}^{soft}(s_{t'},a_{t'})$. Importantly, when $T,\lambda,c_{k} =0$, it returns to the single-step SAC.

\begin{table}[t]
    \centering
    \caption{Choices of different off-policy multi-step algorithms for control. For brevity, we define $\pi_{Q_{k}}$ as the target policy in SAC and denote $\rho_{k}(s,a)$ as $\frac{\pi_{k}(a|s)}{\mu_{k}(a|s)}$.}
    \begin{tabular}{cccc}
    \toprule
        Algorithm & $c_{k} $ & $\pi_{k}$ & Conservative  \\
    \midrule
         Retrace \citep{munos2016safe} & min$\{1,\rho_{k}\}$ & any  & YES\\
         Importance Sampling (IS) & $\rho_{k}$ & any & YES \\
         Tree Backup (TBL) \citep{precup2000eligibility} & $\pi_{k}$ & any &  YES\\
         Q-lambda (PQL) \citep{peng1994incremental} & 1 & $\lambda  \pi_{Q_{k}}+(1-\lambda)\mu_{k}$ & NO \\
         Uncorrected n-step return & 1 & any & NO \\
    \bottomrule
    \end{tabular}
    \label{tab:trace_choices}
\end{table}

Given the choices of $c_{k}$ and $\pi_{k}$ in Table \ref{tab:trace_choices}, we recover a few known algorithms \citep{peng1994incremental,precup2000eligibility,munos2016safe}. Importantly, according to \citep{kozuno2021revisiting}, an algorithm is called \textit{conservative}, if $c_{k}(s,a) \in [0, \pi_{k}(a|s)/ \mu_{k}(a|s)]$ for any $k$ and $(s,a) \in \mathcal{S} \times \mathcal{A}$. See Table \ref{tab:trace_choices} for the classification of algorithms. Through choosing different trace algorithms (\textit{conservative} or \textit{non-conservative}) and controlling the values of $T$ and $\lambda$ in Equation \eqref{retrace}, Trace-SAC is able to handle the trade-off of bias and variance in the off-policy value estimation in the financial environment.

\textbf{Parameter Updates}. As discussed above, large continuous action space drive us to use function approximators for both the Q-function and policy function. To that end, we consider parameterized soft Q-function $Q_{\theta}(s_{t},a_{t})$, a tractable policy distribution $\pi_{\phi}(\cdot |s_{t}) \in [0,1]$, and a policy function  $a_{t} =f_{\phi}(s_{t};\epsilon)$, where $f_{\phi}$ is realized by the reparameterization trick using a network transformation. The parameters of these neural networks are $\theta$ and $\phi$ respectively. In practice, we use the Long Short Term Network (LSTM) \citep{hochreiter1997long} to encoder the stock features for both value and policy approximators. The value function is followed by Fully Connected Layers to output a single Q-value while the policy output mean and standard deviation vectors to parameterize Gaussian distribution with the support in [-h\_max,h\_max]. We will next dive into the update rules for these parameter vectors.

The update rules for Trace-SAC is very similar to the original SAC algorithm. Suppose $\mathcal{D}$ is the replay buffer, the multi-step soft value function is trained to minimize the Least Square Temporal Difference (LSTD):

\begin{equation}
\label{q_loss}
    L_{Q}(\theta) = \mathbb{E}_{(s_{t},a_{t})\sim \mathcal{D}}[\frac{1}{2}(Q_{\theta}(s_{t},a_{t})-\mathcal{R'}_{\lambda}^{c_{k},\mu_{\phi_{k}},\pi_{\phi_{k}}}Q_{\theta_{k}}(s_{t},a_{t}))^{2}],
\end{equation}

where $\mathcal{R'}_{\lambda}^{c_{k},\mu_{\phi_{k}},\pi_{\phi_{k}}}$is the Trace-SAC operator discussed above. Next, the policy parameters can be learned by minimizing:

\begin{equation}
\label{policy_loss}
    L_{\pi}(\phi)=\mathbb{E}_{s_{t} \sim \mathcal{D}, \epsilon_{t} \sim \mathcal{N}(0,1)}[\alpha \log \pi_{\phi}(f_{\phi}(s_{t};\epsilon_{t})|s_{t})-Q_{\theta_{k}}(s_{t},f_{\phi}(s_{t};\epsilon_{t}))].
\end{equation}

Since we disable the self-tuning for the alpha in the original SAC and set it to be a constant term, there is no loss for the alpha function. Finally, putting Equation \eqref{q_loss} and \eqref{policy_loss} together, the update of $\theta$ and $\alpha$ are given by $\theta_{k+1} \leftarrow \theta_{k} - \lambda_{\theta} \nabla_{\theta} L_{Q}(\theta_{k})$ and $\phi_{k+1} \to \phi_{k} - \lambda_{\phi} \nabla_{\phi}L_{\pi}(\phi)$.

Furthermore, Trace-SAC also exploits the double Q-learning structure to mitigate positive bias in the policy improvement step which is known to deteriorate the performance of estimating value function \citep{hasselt2010double}. Generally, Trace-SAC alternates between collecting experience from the financial environment with the current policy and updating the function approximators using behavior policies sampled from a replay buffer. In practice, we allow the Trace-SAC agent to take a single environment step followed by several gradient steps so that it can better capture the patterns in the financial environment. We  also observe that using delayed policy updates in the financial environment will also facilitate the convergence of value function.

\section{Results}

\begin{table}[t]
    \centering
    \caption{Average reward $\pm$ standard deviation for different algorithms on the four different validation environment. Here reward is calculated by $\log(v_{T}/v_{0})$ (unit: \%), where $T$ is the termination state. The highest reward for each environment is highlighted in bold.}
    \begin{tabular}{ccccc}
    \toprule
        & Env0 & Env1 & Env2 & Env3 \\
    \midrule
    Retrace  & 0.0267 $\pm$ 0.1171 & 0.2800 $\pm$ 0.2146 & 0.1043 $\pm$ 0.0782 & 0.3142 $\pm$ 0.4087 \\
    Tree Backup  & 0.0331 $\pm$ 0.0663 & 0.0956 $\pm$ 0.4929 & 0.1228 $\pm$ 0.1300 & 0.5661 $\pm$ 0.5387 \\
    Q($\lambda$) & 0.0422 $\pm$ 0.1105 & \textbf{0.6042 $\pm$ 1.0488} & \textbf{0.1340 $\pm$ 0.1800} & 0.2634 $\pm$ 0.5851 \\
    IS & 0.0271 $\pm$ 0.0605 & 0.1338 $\pm$ 0.6670 & 0.1100 $\pm$ 0.0785 & \textbf{0.7781 $\pm$ 0.5282} \\
    Uncorrect TD($\lambda$) & \textbf{0.0557 $\pm$ 0.0888} & 0.4768 $\pm$ 0.7148 & -0.0869 $\pm$ 0.2538 & 0.2571 $\pm$ 0.5830 \\
    Single-step SAC & 0.0179 $\pm$ 0.0955 & 0.4664 $\pm$  2.5328 & 0.0339 $\pm$ 0.1174 & 0.5691 $\pm$ 1.0788 \\
    Market & -0.5440 & 1.3937 & 0.0425 & 5.9150 \\
    \bottomrule
    \end{tabular}
    \label{tab:rew_table}
\end{table}

\subsection{Evaluations}
We evaluate the overall performance of the proposed Trace-SAC on the four independent environments shown in Figure \ref{fig:ev1}. On each separated environment, we use the cumulative log return $R_{t}$ from start time $t$ to end time $T$ on the validation set as the performance metric. Mathematically, it is given by $R_{t} = \log(v_{T}/v_{0})$, where $v_{T}$ is the total wealth at the termination state and $v_{0}$ is the initial wealth or balance the trading agent holds. Furthermore, our experiment uses the market value and single-step SAC as two benchmarks.

\subsection{Trace Comparison} 
\begin{figure}[!t]
        \centering
        \begin{subfigure}[b]{0.48\textwidth}
            \centering
            \includegraphics[width=\textwidth]{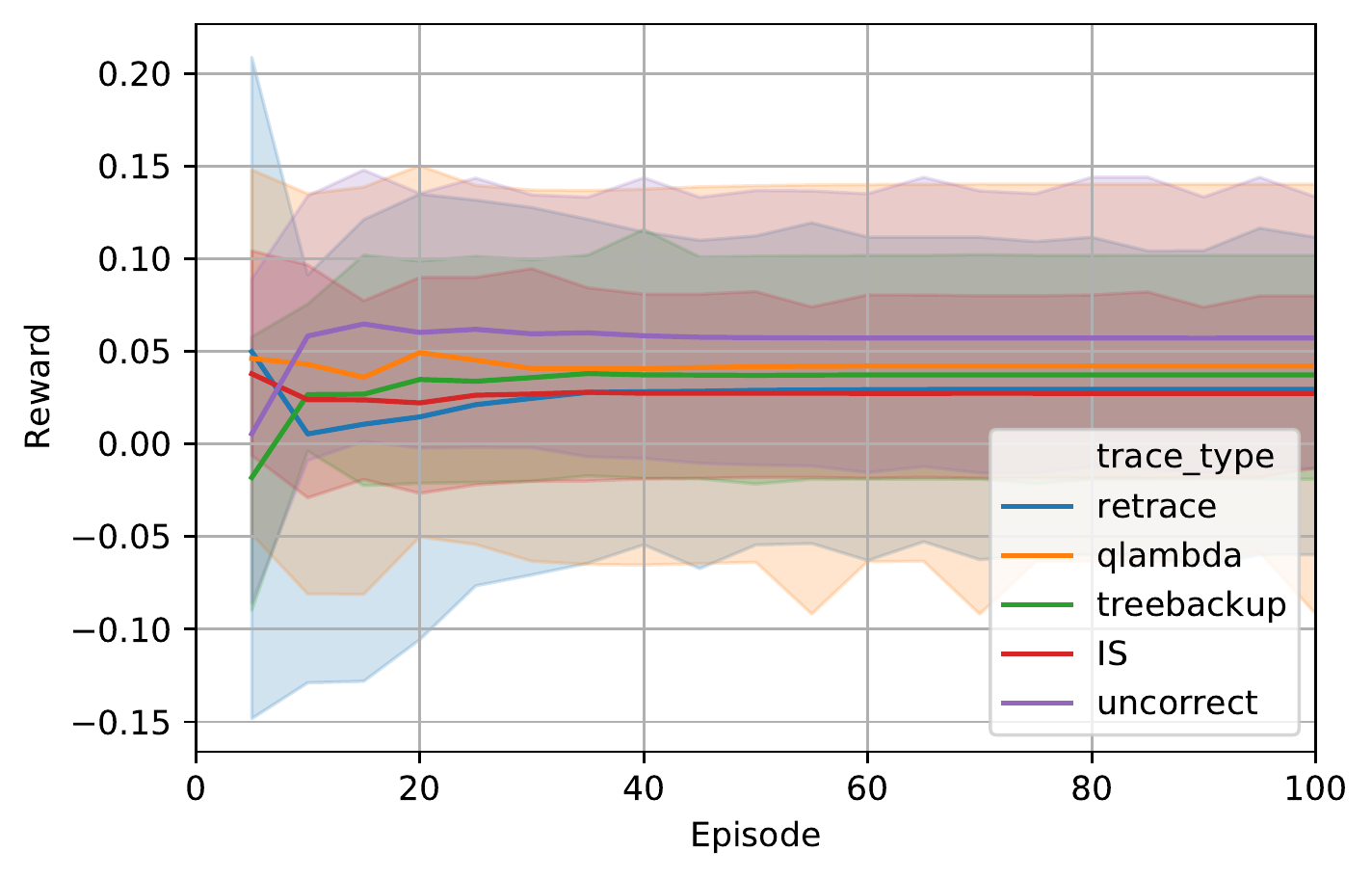}
            \caption[Env No.0]%
            {{\small Env No.0}}    
            \label{fig:1}
        \end{subfigure}
        \hfill
        \begin{subfigure}[b]{0.48\textwidth}  
            \centering 
            \includegraphics[width=\textwidth]{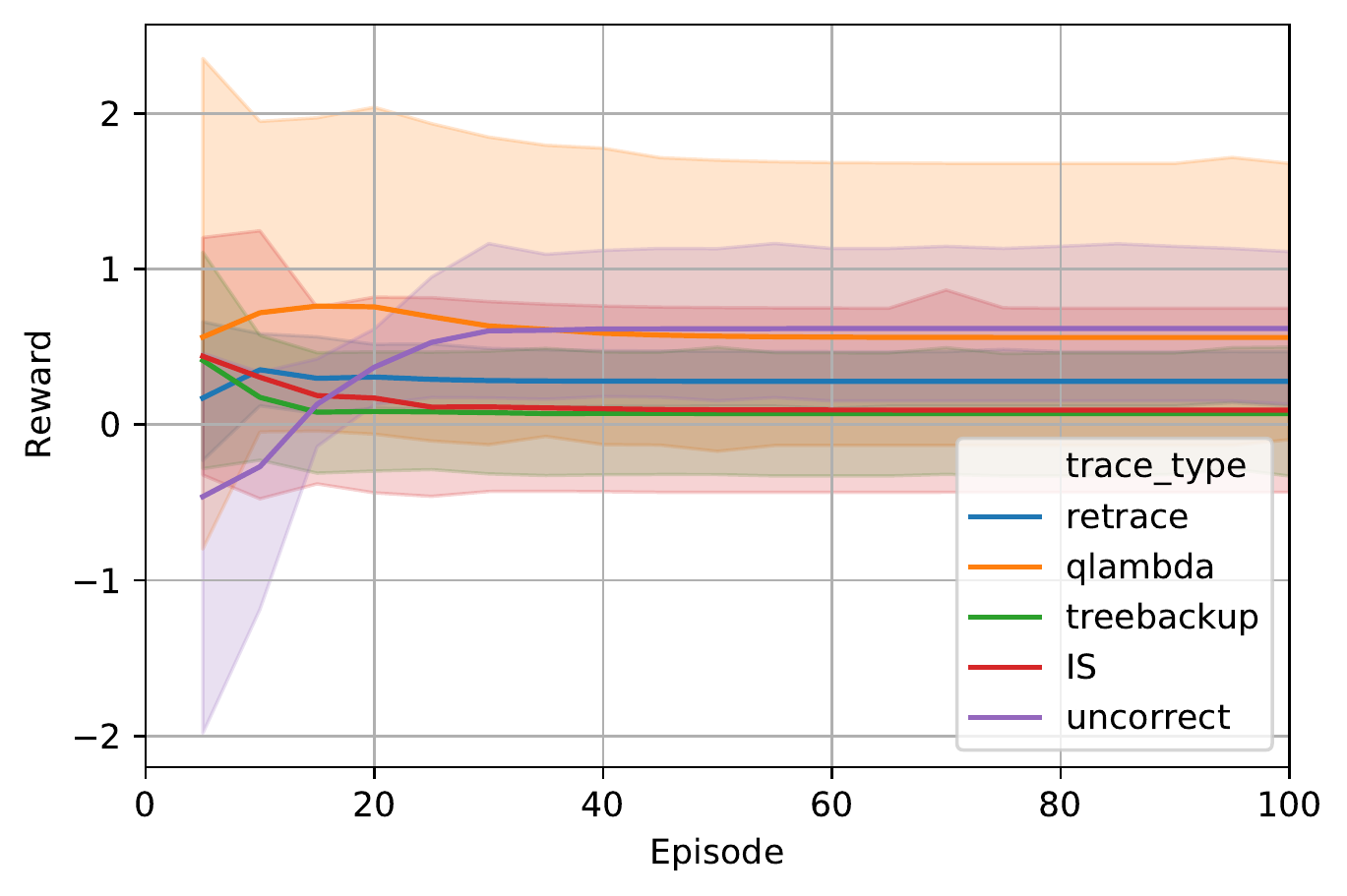}
            \caption[]%
            {{\small Env No.1}}    
            \label{fig:2}
        \end{subfigure}
        \bigskip
        \begin{subfigure}[b]{0.48\textwidth}   
            \centering 
            \includegraphics[width=\textwidth]{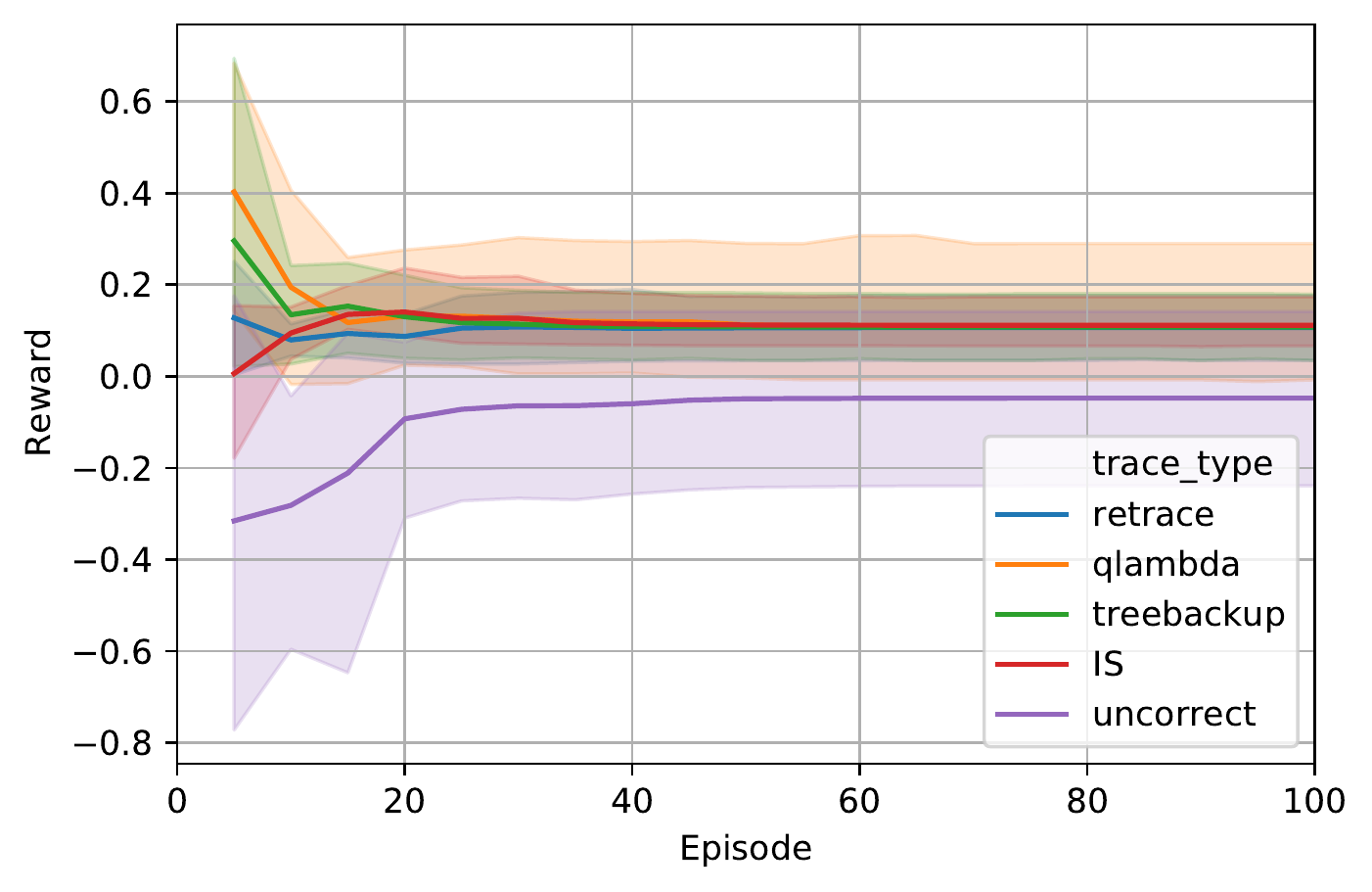}
            \caption[]%
            {{\small Env No.2}}    
            \label{fig:3}
        \end{subfigure}
        \hfill
        \begin{subfigure}[b]{0.48\textwidth}   
            \centering 
            \includegraphics[width=\textwidth]{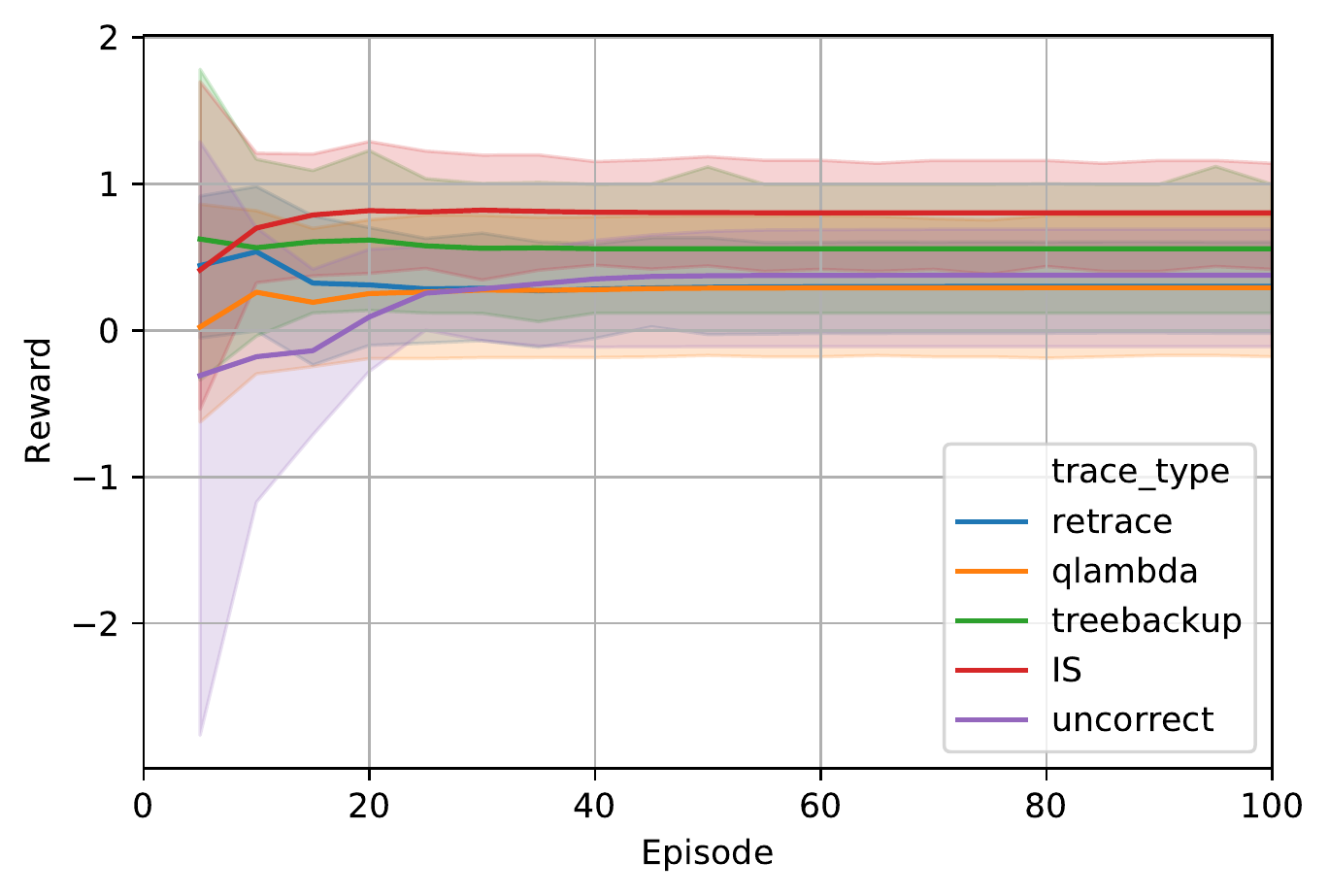}
            \caption[]%
            {{\small Env No.3}}    
            \label{fig:4}
        \end{subfigure}
        \caption[ The average and standard deviation of critical parameters ]
        {\small The validation reward curve with 95 \% CI for different trace algorithms on the four independent environments} 
        \label{fig:trace_env}
    \end{figure}

In Table \ref{tab:rew_table}, we conduct the experiment on five different Trace-SAC algorithms in the four environments. Specifically, we choose (1) Retrace; (2) Tree Backup; (3) Peng's Q($\lambda$) with a fixed $\lambda$; (4) Importance Sampling; (5) Uncorrect n-step with a fixed n to compare with the baselines. Among all the multi-step algorithms, Importance Sampling is conservative and is often the most used algorithm to correct the distribution shift. Retrace is also a representative conservative, while the uncorrected n-step operator is the most commonly used non-conservative operator. The reward curves on the four environments, along with five different algorithms, are shown in Figure \ref{fig:trace_env}.

From Table \ref{tab:rew_table}, we can observe that the five trace algorithms implemented on SAC generally have higher rewards than the single-step SAC in all of the four environments. Interestingly, we discover the single-step SAC has a much higher variance than the multi-step algorithms, especially in Env1 and Env3. Moreover, we observe that the non-conservative algorithms like Q($\lambda$) and uncorrected TD($\lambda$) have remarkably outperformed the conservative ones like IS and Retrace in Env0 and Env1. This observation is consistent with the theoretical finding in \cite{kozuno2021revisiting}, which explains the reason non-conservative algorithms empirically outperform the conservative ones is that the convergence to the optimal policy is recovered when the behavior policy closely tracks the target policy.

\subsection{Environment Comparison}

\begin{figure}
        \centering
        \begin{subfigure}[b]{0.48\textwidth}
            \centering
            \includegraphics[width=\textwidth]{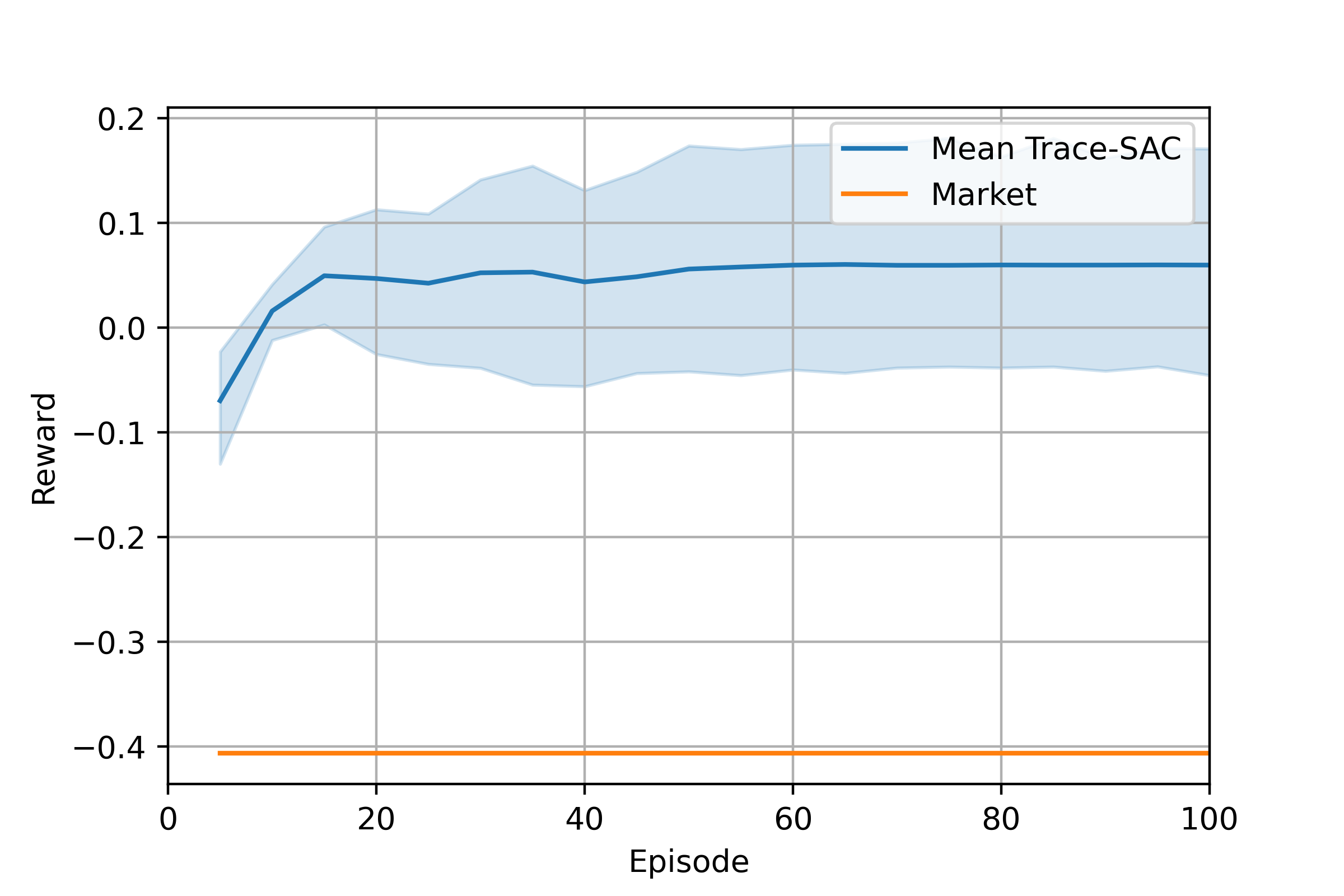}
            \caption[Env No.0]%
            {{\small Env No.0}}    
            \label{fig:1}
        \end{subfigure}
        \hfill
        \begin{subfigure}[b]{0.48\textwidth}  
            \centering 
            \includegraphics[width=\textwidth]{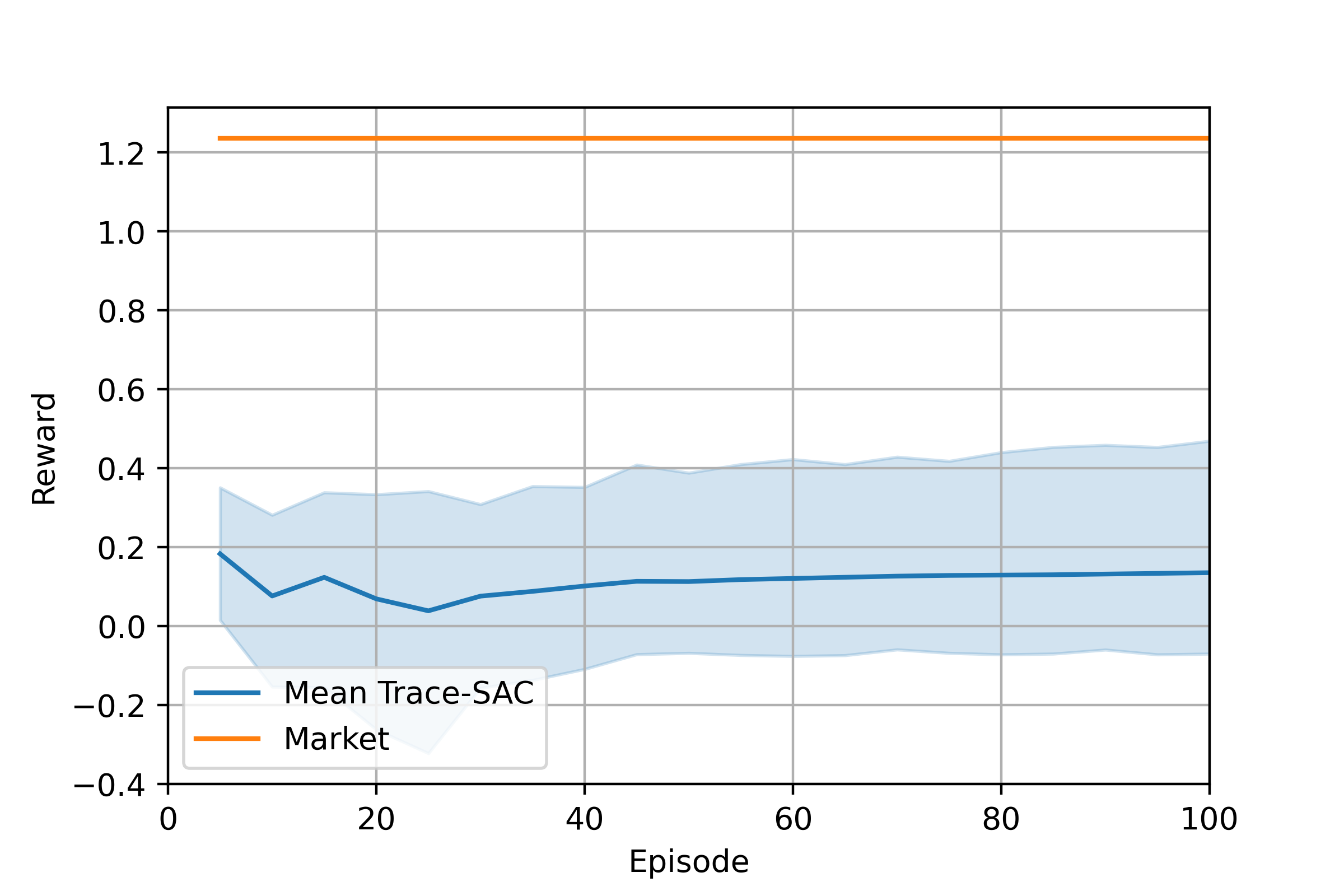}
            \caption[]%
            {{\small Env No.1}}    
            \label{fig:2}
        \end{subfigure}
        \bigskip
        \begin{subfigure}[b]{0.48\textwidth}   
            \centering 
            \includegraphics[width=\textwidth]{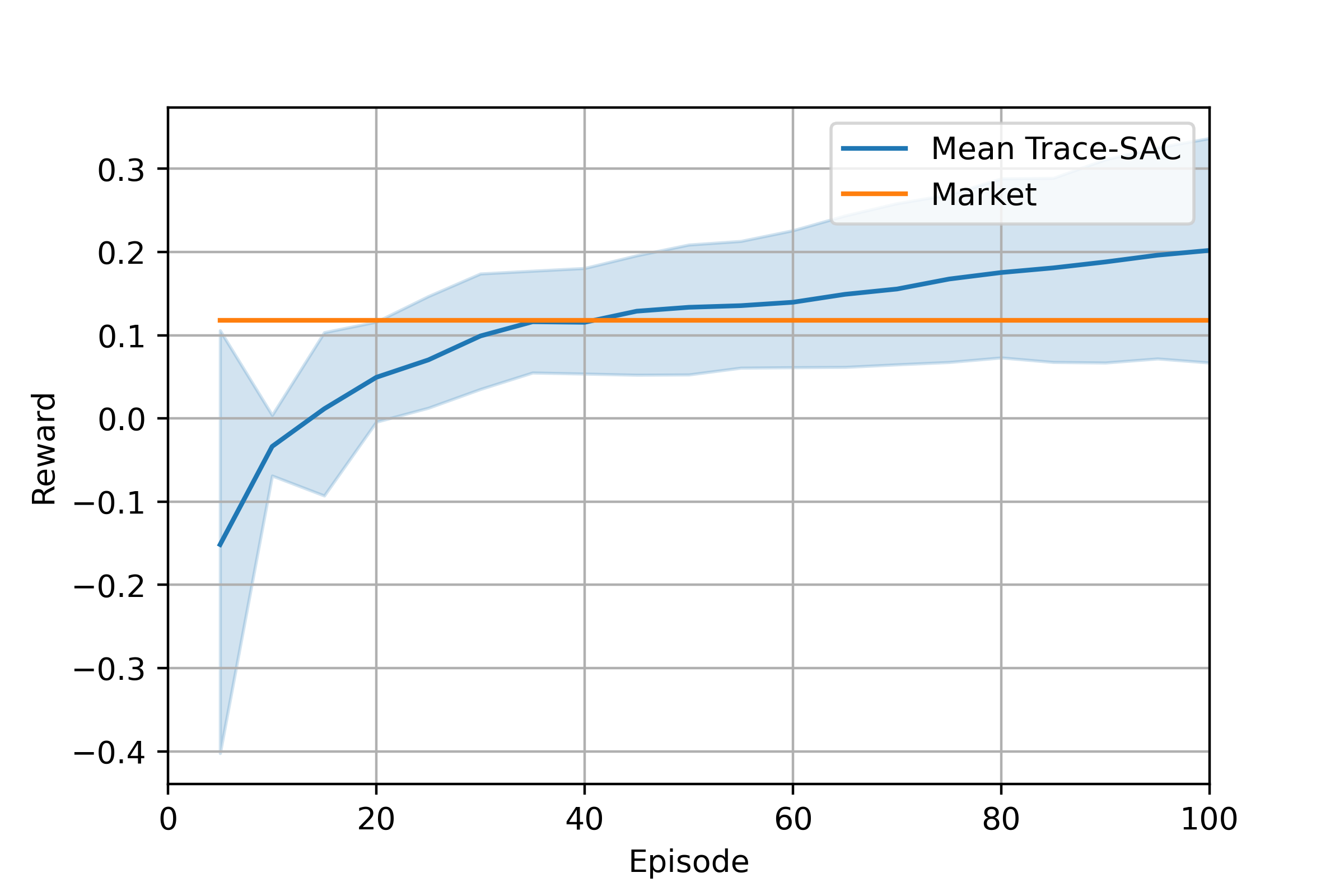}
            \caption[]%
            {{\small Env No.2}}    
            \label{fig:3}
        \end{subfigure}
        \hfill
        \begin{subfigure}[b]{0.48\textwidth}   
            \centering 
            \includegraphics[width=\textwidth]{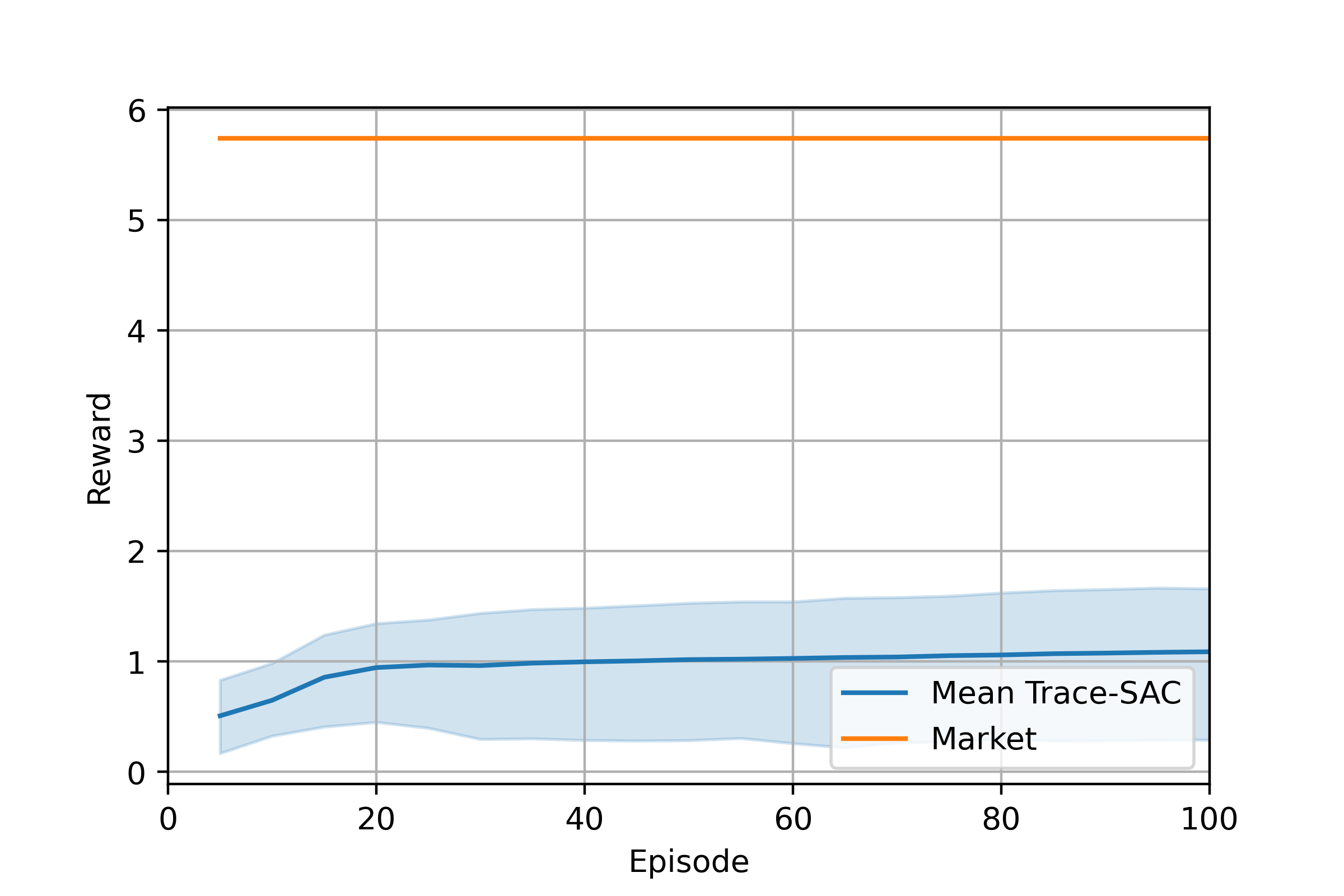}
            \caption[]%
            {{\small Env No.3}}    
            \label{fig:4}
        \end{subfigure}
        \caption[ The average and standard deviation of critical parameters ]
        {\small The validation reward curve of Trace-SAC (using Retrace) with 95\% CI on the four independent environments.} 
        \label{fig:env}
    \end{figure}

In Figure \ref{fig:env}, we train the Trace-SAC using a conservative Retrace algorithm and validate it on the four independent environments. We observe that Trace-SAC has a nice convergence property in all four financial environments. Meanwhile, combined with the information in Table \ref{tab:rew_table}, retrace gives a lower variance than the other four trace algorithms. 

To see the training difference between the four environments, we observe that Trace-SAC has significantly outperformed the market in Env0, and it also has successfully overtaken the market after 40 episodes in Env2. However, retrace has severely underperformed the market in both Env1 and Env3. Combining with the individual environment information in Figure \ref{fig:ev1}, we observe that in Env1, the training environment (red shaded area) is in an active accenting area while the validation environment (yellow and green shaded area) is either in a plateau or in a passive descending area. Similarly, in Env3, the training environment (red shaded area) is in a flat and smooth ascending period, while the validation environment displays a sudden jump from the basin. These discrepancies between the training and validation environment may cause the Trace-SAC agent to fail to capture the trends and underperform the market.

\section{Conclusion}

In this work, we propose Safe-FinRL, which mainly focuses on the challenges of non-stationarity and the bias and variance trade-off when applied to the real financial environment. To address these two problems, firstly, we have separated the long financial time series into the four independent near-stationary financial environments so that we could reduce the negative effect a non-stationary financial environment will bring to the function approximation. Secondly, we have implemented both non-conservative and conservative trace algorithms to estimate the value function in an off-policy financial environment. Empirical, the proposed Safe-FinRL has significantly reduced both the bias and the variance when estimating the value function in a volatile and heterogeneous financial environment and obtained positive returns in all of the proposed environments.

In the future, this work can be extended in three directions. Firstly, more extensive experiments should be conducted on the ablation study of a proper choice of $\lambda$ and step size n. Secondly, the attempt to interpret the trading strategies generated by the DRL agent can also be integrated into the result part. Finally, separating the financial time series manually into small parts may still lead to a non-stationary environment; thus, more sophisticated methods should be developed to handle this problem.

\newpage
\bibliography{main}

\begin{thebibliography}{26}
\providecommand{\natexlab}[1]{#1}
\providecommand{\EM}{\em}
\providecommand{\RNtxt}{\relax}
\RNtxt{}

\bibitem[Fang et~al.(2021)Y.~Fang, K.~Ren, W.~Liu, D.~Zhou, W.~Zhang, J.~Bian,
  Y.~Yu, T.-Y. Liu]{fang2021universal}
{\EM Fang Yuchen, Ren Kan, Liu Weiqing, Zhou Dong, Zhang Weinan, Bian Jiang,
  Yu~Yong, Liu Tie-Yan}.
\newblock Universal Trading for Order Execution with Oracle Policy Distillation
  \allowbreak\newblock// arXiv preprint arXiv:2103.10860. 2021.

\bibitem[Fujimoto et~al.(2018)S.~Fujimoto, H.~Hoof,
  D.~Meger]{fujimoto2018addressing}
{\EM Fujimoto Scott, Hoof Herke, Meger David}.
\newblock Addressing function approximation error in actor-critic methods
  \allowbreak\newblock// International conference on machine learning. 2018.
  1587--1596.

\bibitem[Haarnoja et~al.(2017)T.~Haarnoja, H.~Tang, P.~Abbeel,
  S.~Levine]{haarnoja2017reinforcement}
{\EM Haarnoja Tuomas, Tang Haoran, Abbeel Pieter, Levine Sergey}.
\newblock Reinforcement learning with deep energy-based policies
  \allowbreak\newblock// International Conference on Machine Learning. 2017.
  1352--1361.

\bibitem[Haarnoja et~al.(2018)T.~Haarnoja, A.~Zhou, P.~Abbeel,
  S.~Levine]{haarnoja2018soft}
{\EM Haarnoja Tuomas, Zhou Aurick, Abbeel Pieter, Levine Sergey}.
\newblock Soft actor-critic: Off-policy maximum entropy deep reinforcement
  learning with a stochastic actor \allowbreak\newblock// International
  conference on machine learning. 2018.  1861--1870.

\bibitem[Hasselt(2010)H.~Hasselt]{hasselt2010double}
{\EM Hasselt Hado}.
\newblock Double Q-learning \allowbreak\newblock// Advances in neural
  information processing systems. 2010. 23.

\bibitem[Hochreiter, Schmidhuber(1997)S.~Hochreiter,
  J.~Schmidhuber]{hochreiter1997long}
{\EM Hochreiter Sepp, Schmidhuber J{\"u}rgen}.
\newblock Long short-term memory \allowbreak\newblock// Neural computation.
  1997. 9, 8. 1735--1780.

\bibitem[Jiang et~al.(2017)Z.~Jiang, D.~Xu, J.~Liang]{jiang2017deep}
{\EM Jiang Zhengyao, Xu~Dixing, Liang Jinjun}.
\newblock A deep reinforcement learning framework for the financial portfolio
  management problem \allowbreak\newblock// arXiv preprint arXiv:1706.10059.
  2017.

\bibitem[Kozuno et~al.(2021)T.~Kozuno, Y.~Tang, M.~Rowland, R.~Munos,
  S.~Kapturowski, W.~Dabney, M.~Valko, D.~Abel]{kozuno2021revisiting}
{\EM Kozuno Tadashi, Tang Yunhao, Rowland Mark, Munos R{\'e}mi, Kapturowski
  Steven, Dabney Will, Valko Michal, Abel David}.
\newblock Revisiting Peng’s Q ( $\lambda$ ) for Modern Reinforcement Learning
  \allowbreak\newblock// International Conference on Machine Learning. 2021.
  5794--5804.

\bibitem[Liang et~al.(2018)Z.~Liang, H.~Chen, J.~Zhu, K.~Jiang,
  Y.~Li]{liang2018adversarial}
{\EM Liang Zhipeng, Chen Hao, Zhu Junhao, Jiang Kangkang, Li~Yanran}.
\newblock Adversarial deep reinforcement learning in portfolio management
  \allowbreak\newblock// arXiv preprint arXiv:1808.09940. 2018.

\bibitem[Lillicrap et~al.(2015)T.~P. Lillicrap, J.~J. Hunt, A.~Pritzel,
  N.~Heess, T.~Erez, Y.~Tassa, D.~Silver, D.~Wierstra]{lillicrap2015continuous}
{\EM Lillicrap Timothy~P, Hunt Jonathan~J, Pritzel Alexander, Heess Nicolas,
  Erez Tom, Tassa Yuval, Silver David, Wierstra Daan}.
\newblock Continuous control with deep reinforcement learning
  \allowbreak\newblock// arXiv preprint arXiv:1509.02971. 2015.

\bibitem[Liu et~al.(2020)X.-Y. Liu, H.~Yang, Q.~Chen, R.~Zhang, L.~Yang,
  B.~Xiao, C.~D. Wang]{liu2020finrl}
{\EM Liu Xiao-Yang, Yang Hongyang, Chen Qian, Zhang Runjia, Yang Liuqing, Xiao
  Bowen, Wang Christina~Dan}.
\newblock Finrl: A deep reinforcement learning library for automated stock
  trading in quantitative finance \allowbreak\newblock// arXiv preprint
  arXiv:2011.09607. 2020.

\bibitem[Mnih et~al.(2013)V.~Mnih, K.~Kavukcuoglu, D.~Silver, A.~Graves,
  I.~Antonoglou, D.~Wierstra, M.~Riedmiller]{mnih2013playing}
{\EM Mnih Volodymyr, Kavukcuoglu Koray, Silver David, Graves Alex, Antonoglou
  Ioannis, Wierstra Daan, Riedmiller Martin}.
\newblock Playing atari with deep reinforcement learning \allowbreak\newblock//
  arXiv preprint arXiv:1312.5602. 2013.

\bibitem[Mnih et~al.(2015)V.~Mnih, K.~Kavukcuoglu, D.~Silver, A.~A. Rusu,
  J.~Veness, M.~G. Bellemare, A.~Graves, M.~Riedmiller, A.~K. Fidjeland,
  G.~Ostrovski, et~al.]{mnih2015human}
{\EM Mnih Volodymyr, Kavukcuoglu Koray, Silver David, Rusu Andrei~A, Veness
  Joel, Bellemare Marc~G, Graves Alex, Riedmiller Martin, Fidjeland Andreas~K,
  Ostrovski Georg, others }.
\newblock Human-level control through deep reinforcement learning
  \allowbreak\newblock// nature. 2015. 518, 7540. 529--533.

\bibitem[Munos et~al.(2016)R.~Munos, T.~Stepleton, A.~Harutyunyan,
  M.~Bellemare]{munos2016safe}
{\EM Munos R{\'e}mi, Stepleton Tom, Harutyunyan Anna, Bellemare Marc}.
\newblock Safe and efficient off-policy reinforcement learning
  \allowbreak\newblock// Advances in neural information processing systems.
  2016. 29.

\bibitem[Nachum et~al.(2019{\natexlab{a}})O.~Nachum, Y.~Chow, B.~Dai,
  L.~Li]{nachum2019dualdice}
{\EM Nachum Ofir, Chow Yinlam, Dai Bo, Li~Lihong}.
\newblock Dualdice: Behavior-agnostic estimation of discounted stationary
  distribution corrections \allowbreak\newblock// Advances in Neural
  Information Processing Systems. 2019{\natexlab{a}}. 32.

\bibitem[Nachum et~al.(2019{\natexlab{b}})O.~Nachum, B.~Dai, I.~Kostrikov,
  Y.~Chow, L.~Li, D.~Schuurmans]{nachum2019algaedice}
{\EM Nachum Ofir, Dai Bo, Kostrikov Ilya, Chow Yinlam, Li~Lihong, Schuurmans
  Dale}.
\newblock Algaedice: Policy gradient from arbitrary experience
  \allowbreak\newblock// arXiv preprint arXiv:1912.02074. 2019{\natexlab{b}}.

\bibitem[Peng, Williams(1994)J.~Peng, R.~J. Williams]{peng1994incremental}
{\EM Peng Jing, Williams Ronald~J}.
\newblock Incremental multi-step Q-learning \allowbreak\newblock// Machine
  Learning Proceedings 1994. 1994.  226--232.

\bibitem[Precup(2000)D.~Precup]{precup2000eligibility}
{\EM Precup Doina}.
\newblock Eligibility traces for off-policy policy evaluation
  \allowbreak\newblock// Computer Science Department Faculty Publication
  Series. 2000. ~80.

\bibitem[Robbins, Monro(1951)H.~Robbins, S.~Monro]{robbins1951stochastic}
{\EM Robbins Herbert, Monro Sutton}.
\newblock A stochastic approximation method \allowbreak\newblock// The annals
  of mathematical statistics. 1951.  400--407.

\bibitem[Rowland et~al.(2020)M.~Rowland, W.~Dabney,
  R.~Munos]{rowland2020adaptive}
{\EM Rowland Mark, Dabney Will, Munos R{\'e}mi}.
\newblock Adaptive trade-offs in off-policy learning \allowbreak\newblock//
  International Conference on Artificial Intelligence and Statistics. 2020.
  34--44.

\bibitem[Schulman et~al.(2015{\natexlab{a}})J.~Schulman, S.~Levine, P.~Abbeel,
  M.~Jordan, P.~Moritz]{schulman2015trust}
{\EM Schulman John, Levine Sergey, Abbeel Pieter, Jordan Michael, Moritz
  Philipp}.
\newblock Trust region policy optimization \allowbreak\newblock// International
  conference on machine learning. 2015{\natexlab{a}}.  1889--1897.

\bibitem[Schulman et~al.(2015{\natexlab{b}})J.~Schulman, P.~Moritz, S.~Levine,
  M.~Jordan, P.~Abbeel]{schulman2015high}
{\EM Schulman John, Moritz Philipp, Levine Sergey, Jordan Michael, Abbeel
  Pieter}.
\newblock High-dimensional continuous control using generalized advantage
  estimation \allowbreak\newblock// arXiv preprint arXiv:1506.02438.
  2015{\natexlab{b}}.

\bibitem[Schulman et~al.(2017)J.~Schulman, F.~Wolski, P.~Dhariwal, A.~Radford,
  O.~Klimov]{schulman2017proximal}
{\EM Schulman John, Wolski Filip, Dhariwal Prafulla, Radford Alec, Klimov
  Oleg}.
\newblock Proximal policy optimization algorithms \allowbreak\newblock// arXiv
  preprint arXiv:1707.06347. 2017.

\bibitem[Sutton et~al.(1999)R.~S. Sutton, D.~McAllester, S.~Singh,
  Y.~Mansour]{sutton1999policy}
{\EM Sutton Richard~S, McAllester David, Singh Satinder, Mansour Yishay}.
\newblock Policy gradient methods for reinforcement learning with function
  approximation \allowbreak\newblock// Advances in neural information
  processing systems. 1999. 12.

\bibitem[Wang et~al.(2021)R.~Wang, H.~Wei, B.~An, Z.~Feng,
  J.~Yao]{wang2021deep}
{\EM Wang Rundong, Wei Hongxin, An~Bo, Feng Zhouyan, Yao Jun}.
\newblock Deep Stock Trading: A Hierarchical Reinforcement Learning Framework
  for Portfolio Optimization and Order Execution \allowbreak\newblock// arXiv
  preprint arXiv:2012.12620. 2021.

\bibitem[Watkins, Dayan(1992)C.~J. Watkins, P.~Dayan]{watkins1992q}
{\EM Watkins Christopher~JCH, Dayan Peter}.
\newblock Q-learning \allowbreak\newblock// Machine learning. 1992. 8, 3.
  279--292.

\end{thebibliography}

\end{document}